\documentclass[12pt]{article}
\usepackage{graphicx}
\usepackage{wrapfig}

\newcommand\pubnumber{FMPI-DNP-2017-11-01}
\newcommand\pubdate{\today}

\def\napoli{Comenius Universty, FMPI, Mlynska dolina F2, Bratislava, Slovakia}
\def\support{\footnote{Work supported by Ministry of Education, Science, Research and Sport of Slovakia.}}

\def\Title#1{\begin{center} {\Large #1 } \end{center}}
\def\Author#1{\begin{center}{ \sc #1} \end{center}}
\def\Address#1{\begin{center}{ \it #1} \end{center}}

\newcommand\pubblock{\rightline{\begin{tabular}{l} \pubnumber\\
         \pubdate  \end{tabular}}}
\newenvironment{Abstract}{\begin{quotation}  }{\end{quotation}}
\newenvironment{Presented}{\begin{quotation} \begin{center} 
             PRESENTED AT\end{center}\bigskip 
      \begin{center}\begin{large}}{\end{large}\end{center} \end{quotation}}

\def\beq{\begin{equation}}
\def\eeq#1{\label{#1}\end{equation}}
\def\eeqn{\end{equation}}

\def\beqa{\begin{eqnarray}}
\def\eeqa#1{\label{#1}\end{eqnarray}}
\def\eeqan{\end{eqnarray}}

\let\bar=\overbar

\def\Dslash{\not{\hbox{\kern-4pt $D$}}}
\def\dslash{\not{\hbox{\kern-2pt $\del$}}}

\def\msb{{\bar{\ssstyle M \kern -1pt S}}}

\begin{document}
\begin{titlepage}
\pubblock

\vfill
\Title{Measurement of the particle production properties with the ATLAS Detector}
\vfill
\Author{Stanislav Tokar\support}
\Address{\napoli}
\vfill
\begin{Abstract}
In this contribution, the results on soft hadron physics concerning the underlying event and Bose-Einsten correlations obtained using data collected in proton-proton collisions with the ATLAS experiment are presented.
\end{Abstract}
\vfill
\begin{Presented}
EDS Blois 2017, Prague, \\ Czech Republic, June 26-30, 2017
\end{Presented}
\vfill
\end{titlepage}
\def\thefootnote{\fnsymbol{footnote}}
\setcounter{footnote}{0}

\section{Introduction}
Experiments at hadron colliders are usually aimed at study of deep inelastic  processes connected with tests of the Standard model (SM) or searches for a new physics. But the deep inelastic studies or the searches  need not only a good understanding of the primary short-distance hard scattering process, but also understanding of the accompanying interactions of the rest of the proton--proton ($pp$) collision, usually called the underlying event (UE).  In addition, this contribution deals also with two particles Bose-Einstein correlations (BEC). 
Presence of BEC is another manifestation of the soft hadron physics that can provide us with the space-time characteristics of hadronization process. 
In this contribution results on the  UE and BEC effects using data collected with the ATLAS detector in $pp$ collisions at centre-of-mass-energies $\sqrt{s}$ = 0.9, 7 TeV (BEC) and 13 TeV (UE) are presented.
\section{Underlying event in $\sqrt{s}$ = 13~TeV $pp$ collisions}

Main underlying event sources are: initial- and final-state radiation (ISR, FSR),
QCD evolution of colour connections between the parton hard scattering and the beam-proton remnants and
 additional hard scatters in the $pp$ collision (multiple partonic interactions (MPI)).
 
The measurement of UE observables with the ATLAS detector \cite{atl_det} was carried out
using charged particles in 1.6 nb$^{-1}$ of $pp$ collisions at 
$\sqrt{s}$ = 13~TeV \cite{atl_UE_2017}.
The UE observables have been  constructed from primary charged particles in the pseudorapidity
range $|\eta | < $ 2.5 and with particle transverse momentum $p_\mathrm{T} >$ 500 MeV. 
For the measurement of UE observables, the azimuthal plane of the event is segmented into several distinct regions with differing sensitivities to the UE. 
The azimuthal angular difference with respect to the leading (highest-$p_\mathrm{T} $) charged
particle, $\Delta \phi =\ \mid \phi - \phi_\mathrm{lead}\mid $, is used to define the regions:
 $\Delta \phi < 60^{\circ} $, the "towards region";
 $60^{\circ} < \Delta \phi < 120^{\circ}$, the "transverse region"; and
 $\Delta \phi  > 120^{\circ}$, the "away region".
The towards and away regions are dominated by particle production from the hard process and are hence relatively insensitive to the softer UE. 
In contrast, the transverse region is more sensitive to the UE, and observables defined inside it are the primary focus of UE measurements. The observables used to study the UE, which are summarized in Table \ref{tab:UE_obs}, are investigated with respect to the leading particle transverse momentum, $p\mathrm{_{T}^{lead}} $, number of charged particles in the transverse region, $N_\mathrm{ch}$, and azimuthal angle from the leading charged particle, $\Delta \phi $.
The {\sc Pythia} 8 \cite{pythia1, pythia2}, {\sc Herwig} 7 \cite{herwig7} and {\sc Epos} \cite{epos1, epos2} Monte Carlo (MC) event
generator models with different tunes are used either for data corrections or for comparison to the final
corrected data distributions.\\
{\bf Unfolding to particle level.} The measured UE distributions were unfolded to 
a particle level, i.e. the reconstructed observables were corrected for detector effects. 
\begin{table}[t]
 \begin{center}
 \begin{tabular}{ll}  
\hline
 Symbol & Description \\ \hline
$<N_\mathrm{ch}/\delta \eta \delta \phi> $ & Mean number of charged particles per unit $\eta - \phi $ (in radians) \\
$<\Sigma p_\mathrm{T}/\delta \eta \delta \phi> $ & Mean scalar $p_\mathrm{T}$ sum of charged particles per unit $\eta - \phi $(in radians)\\
$<\mathrm{mean}\, p_\mathrm{T}> $ & Mean per-event average $p_\mathrm{T}$ of charged particles ($\geq $ 1 charged \\                                  & particle required)\\
 \hline                
 \end{tabular}
 \caption{Definition of the measured observables. 
}
\vspace{-6mm}
 \label{tab:UE_obs}
 \end{center}
 \end{table}

\noindent Practically,  a weighting procedure was applied to take into account inefficiencies due to the trigger selection, vertexing, and track reconstruction. Additionally, also a correction of azimuthal re-orientation of the event, based on HBOM method \cite{hbom}, was applied. 
This correction is connected  with the fact that  the leading charged particle can be reconstructed incorrectly and thereby identification of the towards, transverse, and away regions can differ from that at particle level.

{\bf Systematic uncertainties.} The following sources of systematic uncertainties were taken into account: trigger and vertexing,  track reconstruction - 
mainly from imperfect knowledge of the material in the inner detector,
non-primary particles - from modification of track weights obtained using variations of the fit range in the tail of impact parameter distributions for different MC generators,
 and unfolding procedure - uncertainties associated with the HBOM unfolding at event azimuthal re-orientation correction. 
All these uncertainties are at the level below 1\% and are added quadratically to get the total systematic uncertainty - see details in \cite{atl_UE_2017}. 
 
\begin{figure*}[htb]
	\centering
	\begin{tabular}{ccc}
        \includegraphics[width=0.32\textwidth]{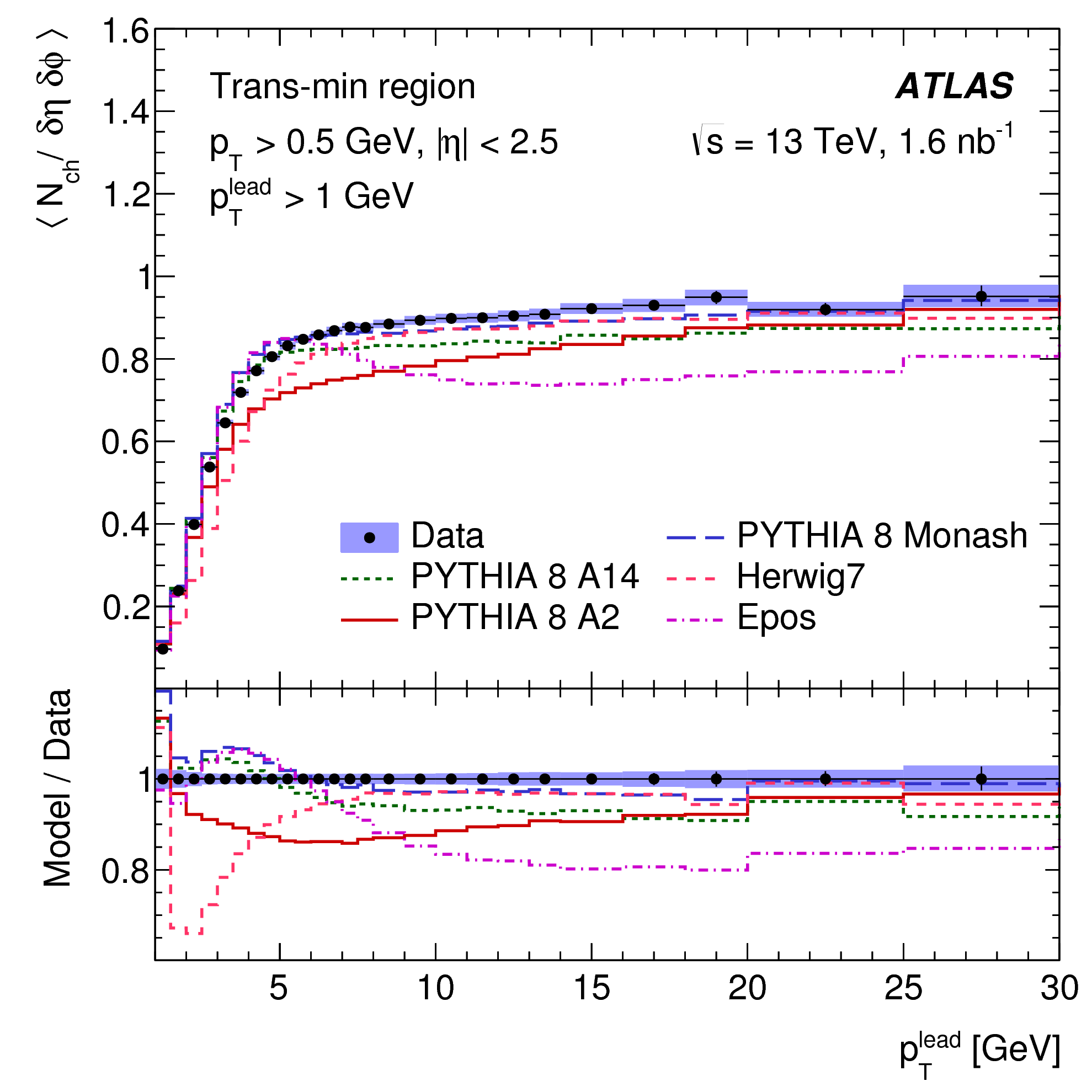} &
        \includegraphics[width=0.32\textwidth]{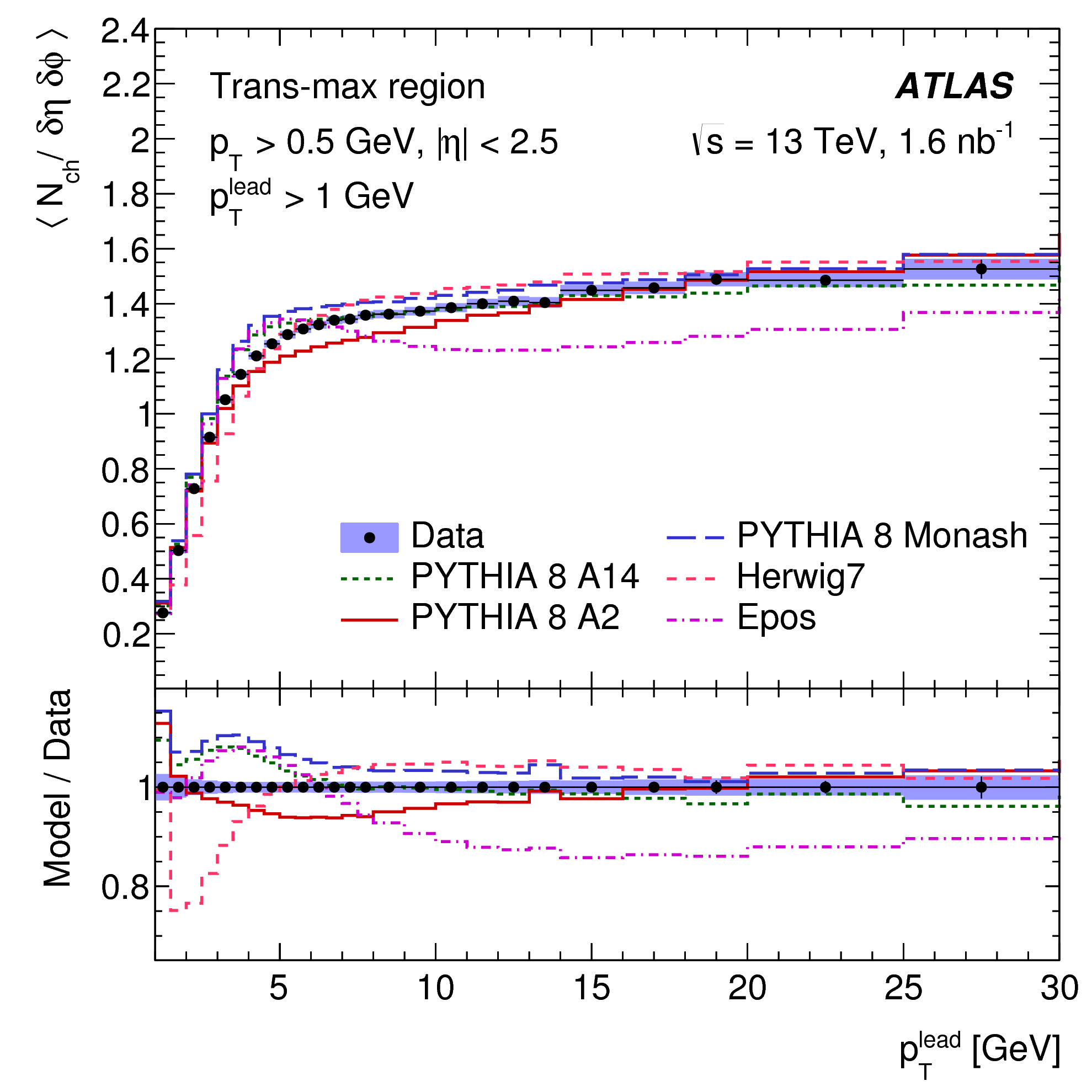} &
        \includegraphics[width=0.32\textwidth]{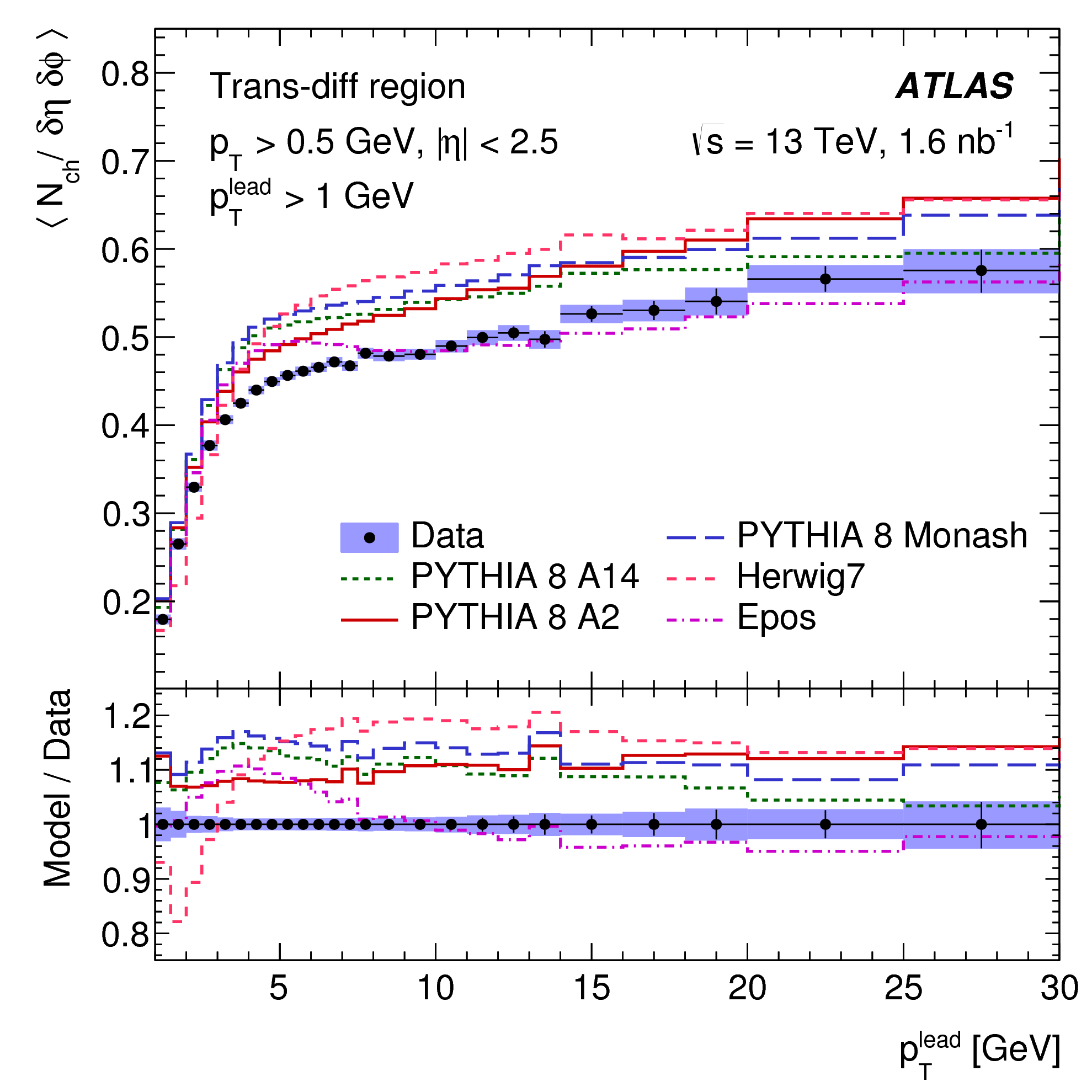} \\
        \includegraphics[width=0.32\textwidth]{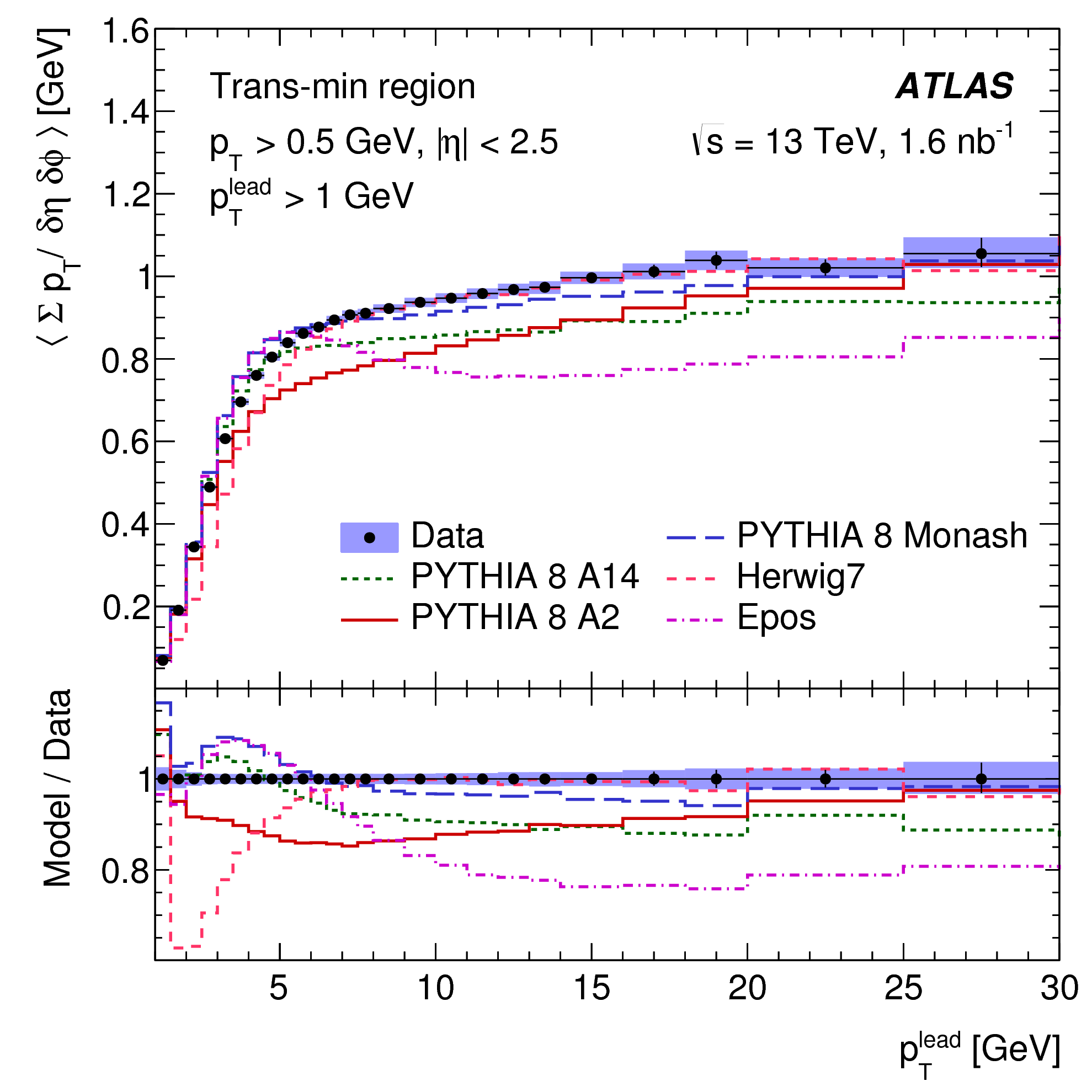} &
        \includegraphics[width=0.32\textwidth]{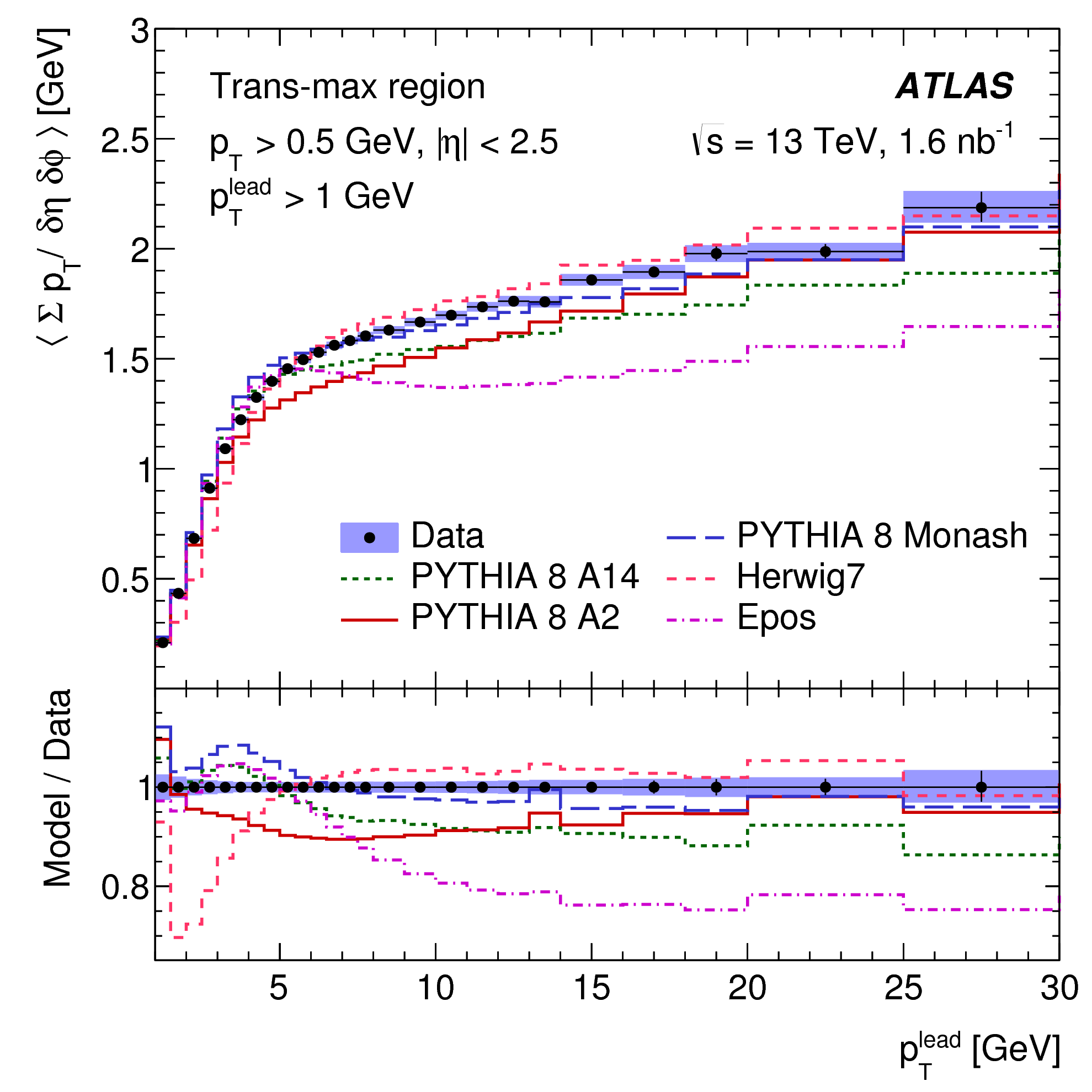} &
        \includegraphics[width=0.32\textwidth]{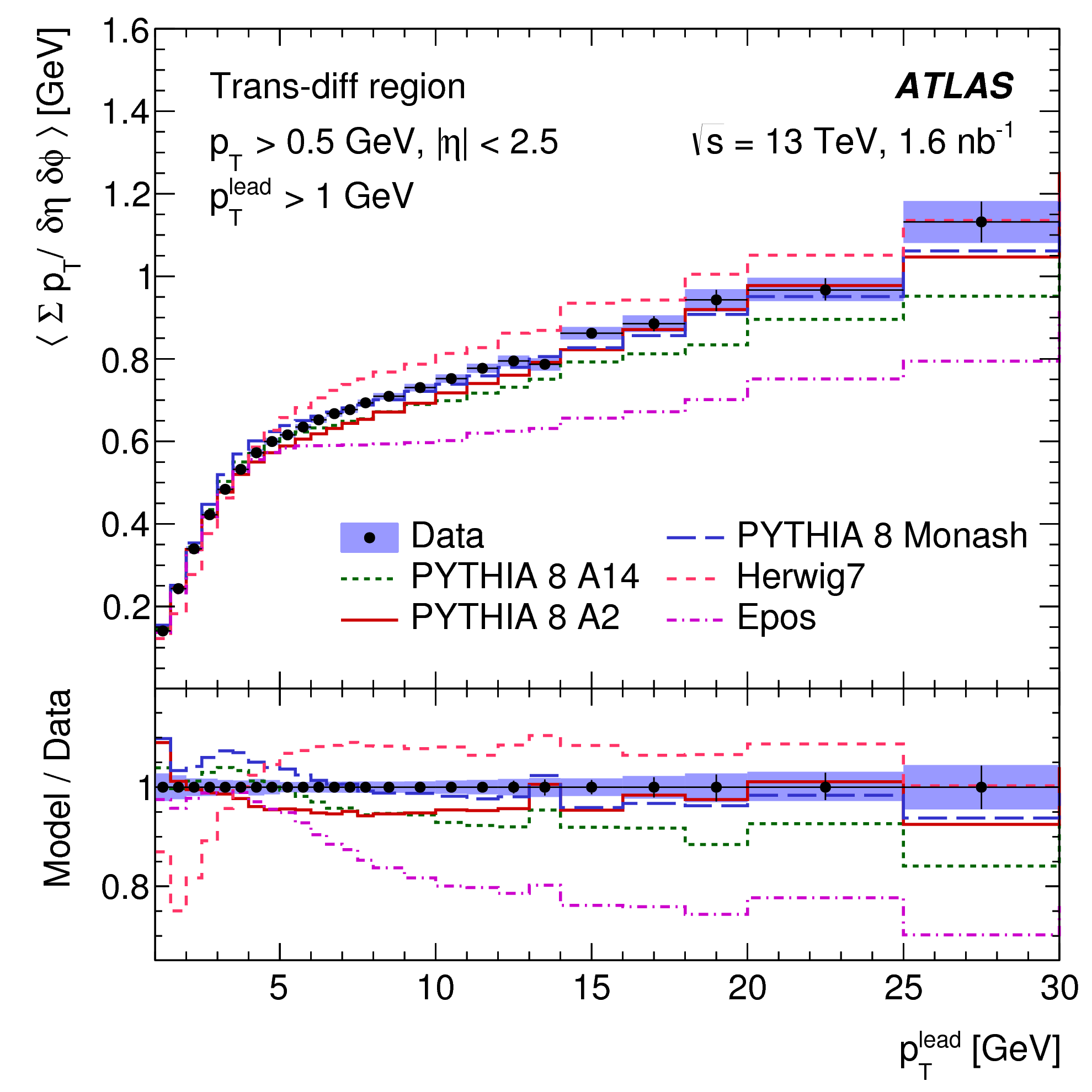} \\
	\end{tabular}
	\caption{Mean densities of charged-particle multiplicity $N_\mathrm{ch}$ (top row) and $\Sigma p_\mathrm{T}$ (bottom row) as a function of $p\mathrm{^{lead}_{T}}$, in the trans-min (left), trans-max (middle) and trans-diff (right) azimuthal regions. The error bars represent statistical uncertainty and the blue band the total combined statistical and systematic uncertainty.}
	\label{fig:UE_manif}
\end{figure*}

 {\bf Results.} Fig. \ref{fig:UE_manif} shows the mean densities of charged-particle multiplicity $N_\mathrm{ch}$ (top row) and $\Sigma p_\mathrm{T}$ (bottom row) as a function of leading charged-particle $p_\mathrm{T}$ in the UE-dominated transverse
region, and its per-event sub-regions trans-min, -max, and -diff. The trans-min is the most sensitive to MPI effects, while the trans-max includes both MPI and hard process contaminations. Hence, the trans-diff is the clearest measure of the hard process  contaminations. 
The comparisons to MC models are also shown in these plots. The best performance comes from the {\sc Pythia} 8 Monash and A2 tunes, with {\sc Herwig} 7 and {\sc Pythia} 8 A14 both wrong by 5–-10\%.
Study of the mean densities of $N_\mathrm{ch}$ and $\Sigma p_\mathrm{T}$ as a function of $p\mathrm{^{lead}_{T}}$ shows an increase in UE activity of approximately 20\% when going from 7~TeV to 13~TeV $pp$ collisions.

\section{Bose-Einstein correlations.}
The Bose-Einstein correlations (BEC) were studied by the ATLAS experiment in $pp$ collisions at $\sqrt{s}$ = 0.9 and 7~TeV \cite{atl_BEC}. 
From theoretical point of view the BEC effect corresponds to an enhancement in two identical boson correlation function when the two particles are near in momentum space. 
They represent a sensitive probe of the space–time geometry of the hadronization region and allow determination of the  size and the shape of the source from which particles are emitted. In general two-particle correlation function is defined as a ratio of the probability to observe simultaneously two particles with four-momenta $p_{1}$ and $p_{2}$ and a product of two one-particle distributions ($C_{2}(p_{1},p_{2})=P\left(p_{1},p_{2}\right)/P(p_{1})P(p_{2})$).

The BEC effect is usually described by a function with two parameters: the effective radius parameter $R$ and the strength
parameter $\lambda $. In this study two parametrizations are used for the correlation $C_{2}$ function:
\begin{equation}
C_{2}^\mathrm{G}(Q)= C_{0}\left(1+\lambda e^{-R^{2}Q^2}\right)(1+\epsilon Q),\ C_{2}^\mathrm{E}(Q)= C_{0}\left(1+\lambda e^{-RQ}\right)(1+\epsilon Q),
\label{C2par}
\end{equation}
where $Q^{2}=-(p_1-p_2)^{2}$ is the Lorentz invariant four-momentum difference squared of the two particles squared. The parametrization with the Gaussian and exponential form of $C_2$ function, which are commonly used, are used also in this study.
From experimental point of view the correlation $C_2$ fuction is defined as a ratio of a signal distribution ($N^\mathrm{LS}$) containing the BEC effect and a reference distribution ($N^\mathrm{ref}$) which does not contain it: $C_{2}(Q)=  N^\mathrm{LS}(Q) / N^\mathrm{ref}(Q)$.
The signal distribution should be created by  pairs of identical particles (like-sign pairs) while the reference distribution should not contain effect of identical particles -- can be created of unlike-sign pairs or artificial distribution (event mixing, opposite hemisphere, ...), see details in Ref. \cite{atl_BEC}. In this ATLAS study, instead of the $C_2$ function so called double ratio $R_2(Q)$ is used:  
\begin{equation}
R_{2}(Q )=  C_{2}^\mathrm{Data}(Q) / C_{2}^\mathrm{MC}(Q),
\label{R2}
\end{equation}
The $R_2$ ratio eliminates problems with energy-momentum conservation, topology, etc. in reference distributions. In addition $C_{2}^\mathrm{MC}$ does not contain BEC effect but it should contain all other correlations present in $C_{2}^\mathrm{Data}$.

{\bf Event and object selection.}
Data are selected with a minimum bias trigger ($\geq $ 1 minimum-bias trigger scintillators hits). In addition, there are requirements on primary vertex selection and on track selection:
each event was required to contain a primary vertex reconstructed from at least two tracks with $p_\mathrm{T} >$ 100 MeV; 
the primary vertex was identified as that with the highest $\Sigma p_\mathrm{T}^{2}$ of its associated tracks;
events containing more than 1 primary vertex with at least 4 associated tracks were removed.
Each track was required to have  hits in both the pixel system and the silicon microstrip detector and  
 in the innermost pixel layer.
 
{\bf Track pair $Q$-distribution correction.}  
Due to inefficiencies in trigger and vertex reconstruction, the measured track pair $Q$-distribution, $N_\mathrm{meas}(Q)$, should be corrected -- the true $Q$-distribution reads $N(Q) = N_\mathrm{meas}(Q)/\left(\epsilon_\mathrm{trig}\epsilon_\mathrm{vert} \right)$, where $\epsilon_\mathrm{trig}$ ($\epsilon_\mathrm{vert}$) is the trigger (vertex) efficiency.
In addition, due to Coulomb interaction in the final state, we need to remove the Coulomb effect from the measured $N_\mathrm{meas}(Q)$:
$N_\mathrm{meas}(Q) =  G(Q)\cdot N(Q),\ G(Q)=2\pi \kappa  /\left(e^{2\pi \kappa  }-1\right)$
where $\kappa $ is the Sommerfeld parameter; $\kappa  >$ 0 ($\kappa  <$ 0) for like-sign (unlike-sign) pairs \cite{atl_UE_2017}.

{\bf Systematic uncertainties.}
The systematic uncertainties were studied  for the parameters $R$ and $\lambda $ of double-ratio correlation function $R_2(Q)$ (see ref.~\cite{atl_UE_2017}). The exponetial fit, which gives better results, was studied. 
The  BEC studies were performed for the full kinematic region at $\sqrt{s}$ = 0.9 and 7 TeV for the minimum-bias and high-multiplicity (HM) events. The total systematic uncertainty for the hadronization radius $R$ (factor $\lambda $) is 13\% (14.8\%) at 0.9~TeV and 10.7\% (9.6\%) at 7~TeV.

{\bf Results.} 
The output of the $R_2$ correlation function analysis are the parameters 
\begin{figure*}[htb]
     \vspace{-3mm}
	\centering
	\begin{tabular}{ccc}
        \includegraphics[width=0.32\textwidth]{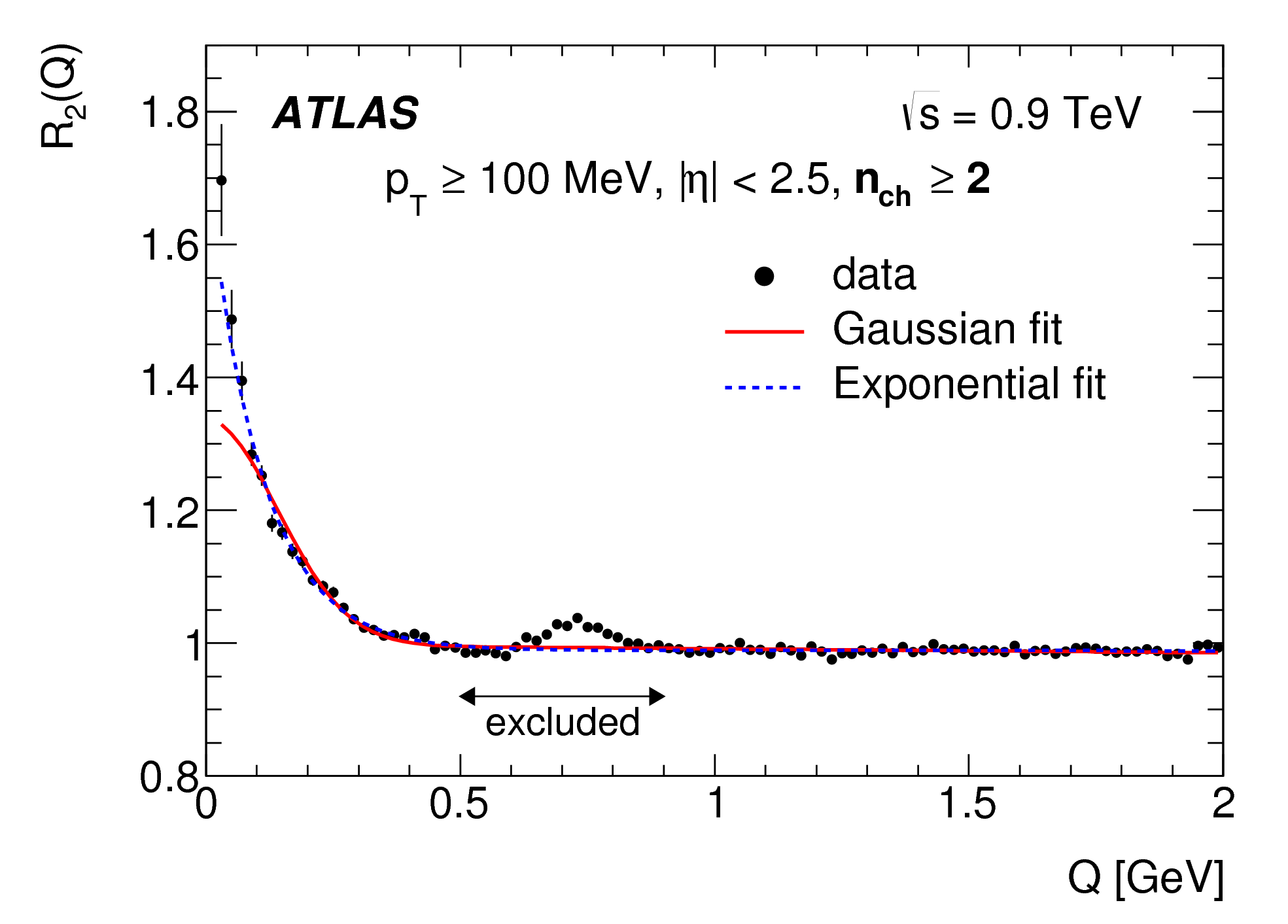} &
        \includegraphics[width=0.32\textwidth]{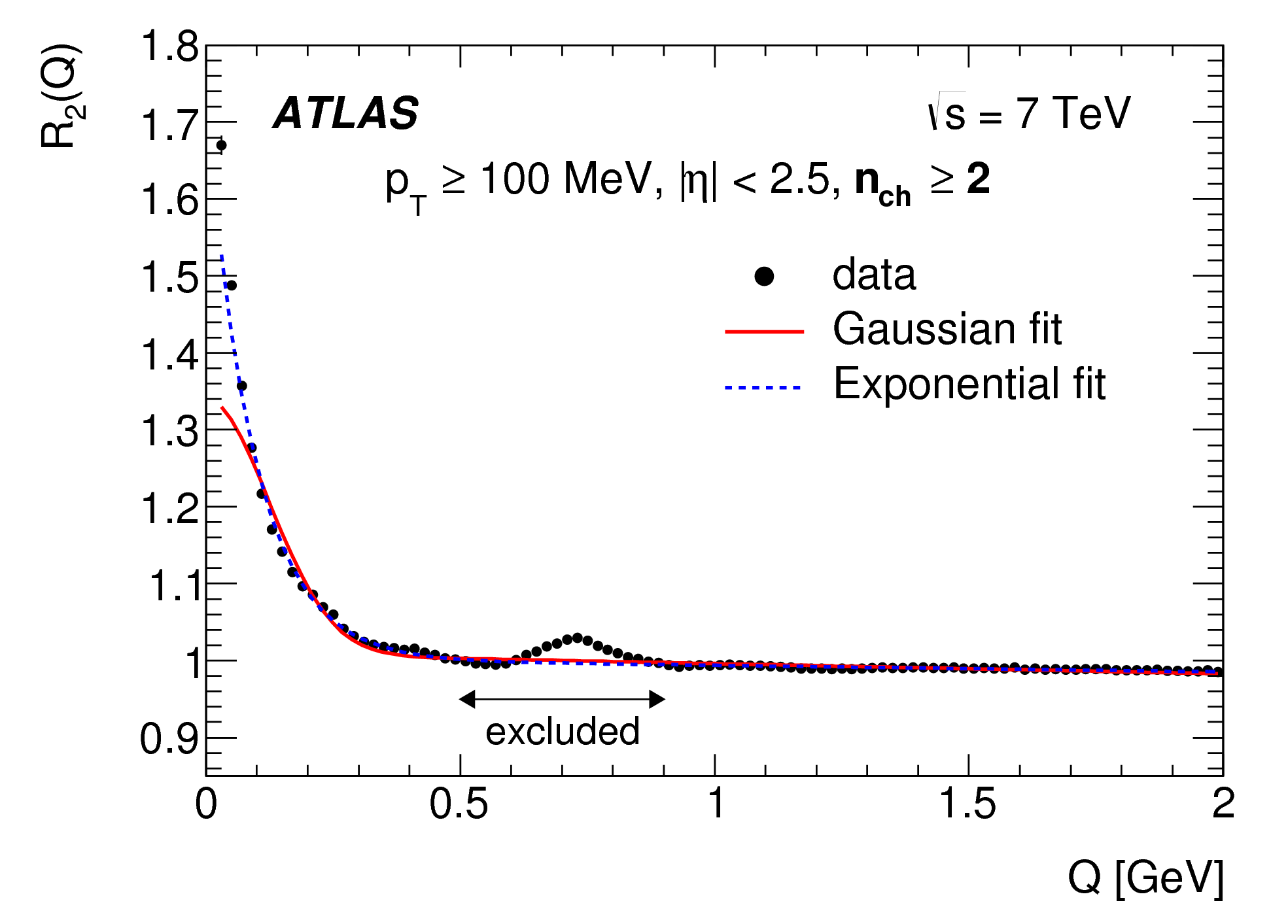} &
        \includegraphics[width=0.32\textwidth]{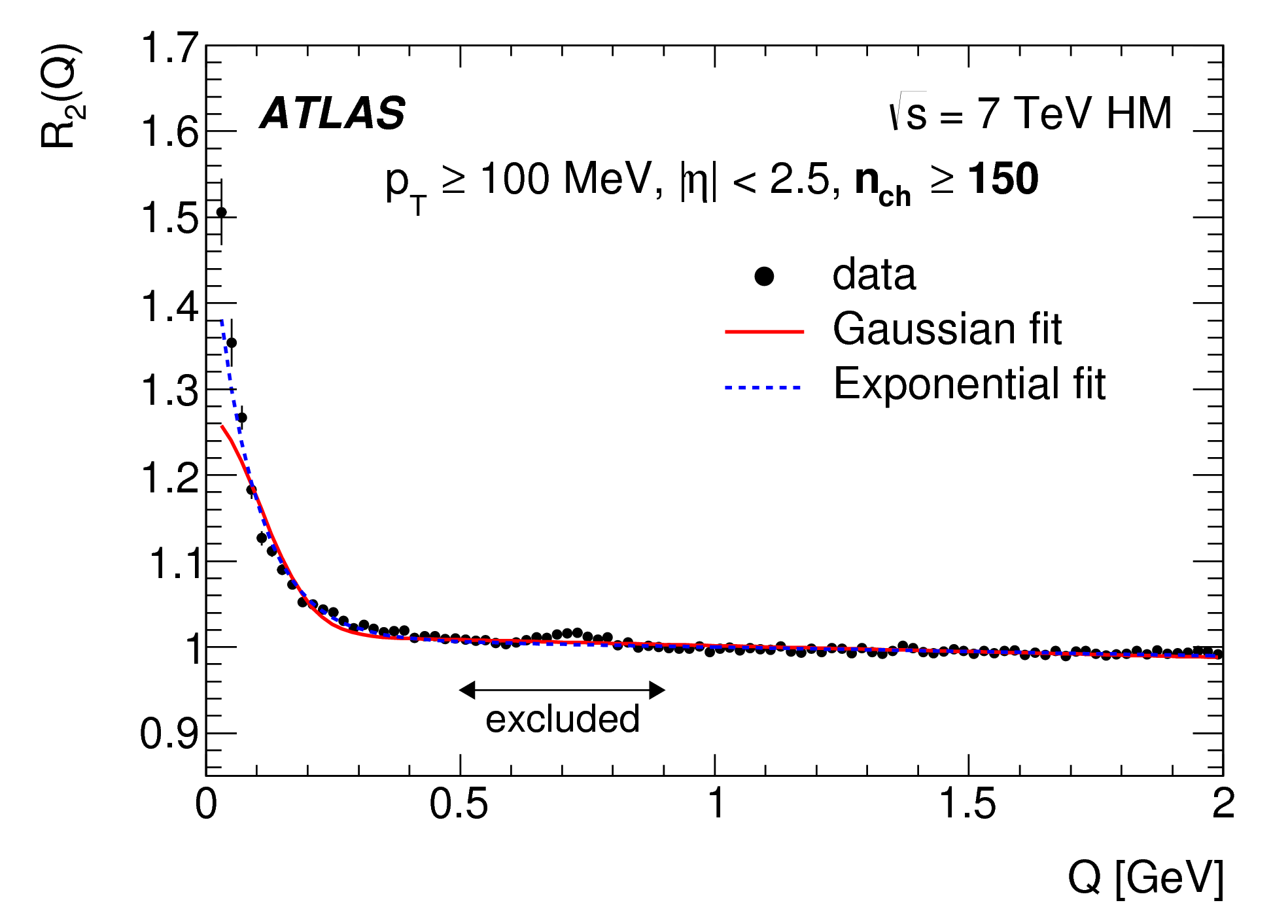} \\
 	\end{tabular}
     \vspace{-4mm}
	\caption{\small Correlation $R_2(Q)$ function for data taken  at $\sqrt{s}$ = 0.9 TeV (left), 7 TeV (midle) and 7 TeV HM (right).}
	\label{fig:BEC_R2}
    \vspace{-3mm}
\end{figure*}
\noindent $R$  (hadronization radius) and $\lambda $ (incoherence factor). Much better description was obtained for the exponential fit. Fig. \ref{fig:BEC_R2} shows the $R_2$ functions reconstructed at $\sqrt{s}$ = 0.9, 7~TeV and 7~TeV with the HM trigger. The data are fitted with Gaussian and exponential fits. 
From Fig.~\ref{fig:BEC_R2} it is clear that the data are much better described by the exponential fit. The bump in resonance region is due to MC overestimation of resonances 
\begin{wrapfigure}{r}{0.40\textwidth}
\vspace*{-3mm}
\centering
\begin{tabular}{c}
\includegraphics[width=0.39\textwidth]{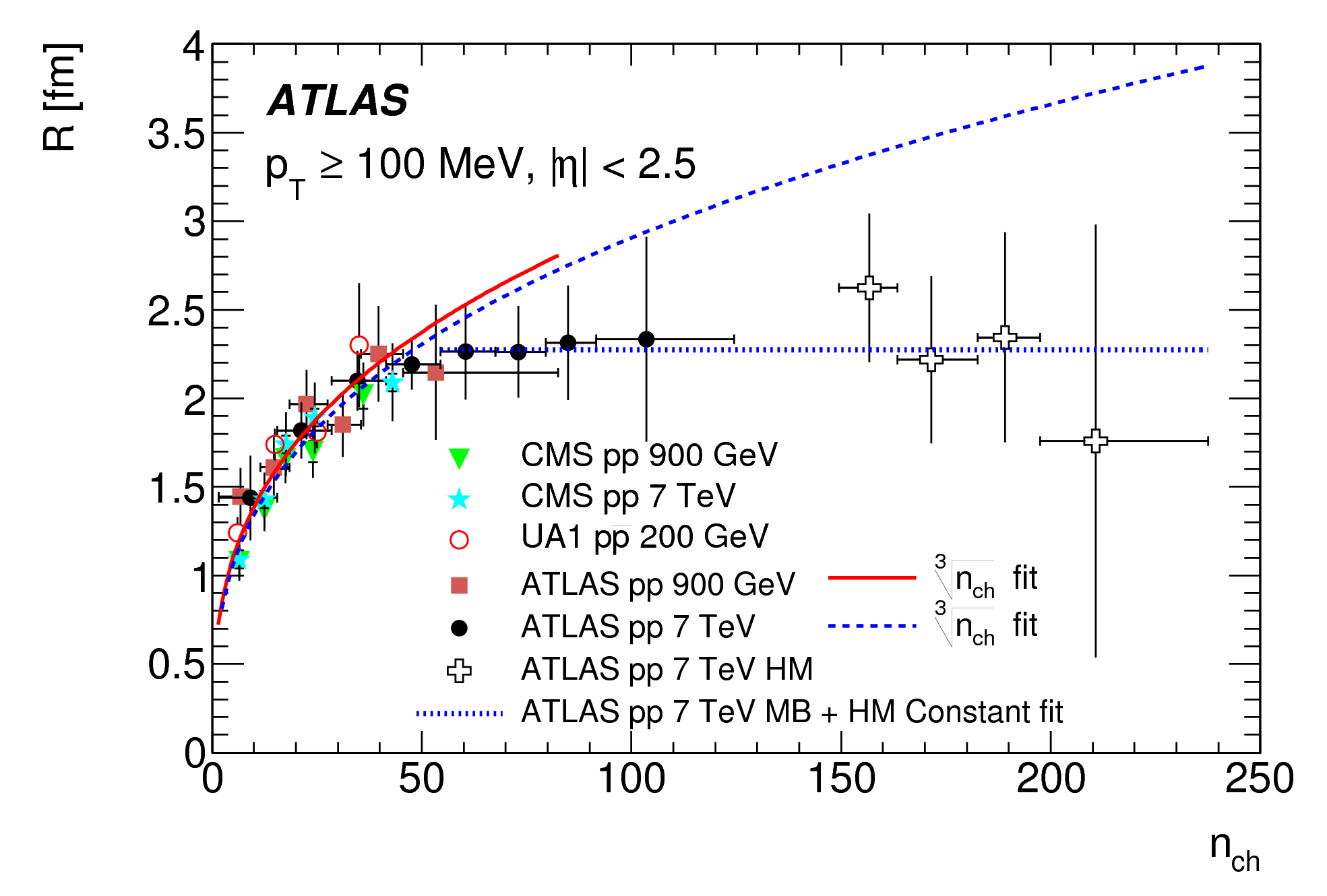}
\end{tabular}
\vspace*{-6mm}
{\small
\caption{Hadronization radius ($R$) vs. multiplicity ($n_\mathrm{ch}$).}
}
\label{fig:R_nch}
\vspace*{-7mm}
\end{wrapfigure}
\noindent (mainly $\rho \rightarrow \pi \pi $).
The region 0.5 –- 0.9~GeV was excluded from the fit. The obtained values of the parameters $R$ and $\lambda $ are the following:\\
  $R$ = 1.83 $\pm $ 0.25,  $\lambda $ = (0.74 $\pm $ 0.11) fm \\ at $\sqrt{s}$ = 0.9 TeV for $n_\mathrm{ch} \geq $ 2\\
  $R$ = 2.06 $\pm $ 0.22,  $\lambda $ = (0.71 $\pm $ 0.07) fm \\ at $\sqrt{s}$ = 7 TeV   for $n_\mathrm{ch} \geq $ 2\\
  $R$ = 3.36 $\pm $ 0.30,  $\lambda $ = (0.74 $\pm $ 0.11) fm \\ at $\sqrt{s}$ = 7 TeV   for $n_\mathrm{ch} \geq $ 150.\\
Multiplicity dependence of the BEC parameters was investigated 
and a saturation
effect for the source size $R$ is seen for multiplicity $n_\mathrm{ch} >$ 55 -- see Fig.~\ref{fig:R_nch}. The BEC parameters dependences on track pair transverse momentum and on particle $p_\mathrm{T}$ were also studied.

\section{Conclusions}
Several distributions  sensitive to UE measured by the ATLAS experiment at 13 TeV $pp$ collisions are presented.
An improvement upon previous ATLAS measurements of UE, performed at 0.9 and 7 TeV
, are achieved.
An increase in UE activity by $\approx $20\% is observed when going from 7 TeV to 13 TeV $pp$ collisions.
MC generators: for most observables the models describe the UE data to a better than 5\% accuracy, but it is greater than the experimental uncertainty.
BEC of the pairs of identical charged particles were measured within $|\eta | <$ 2.5 and $p_\mathrm{T} >$ 100 MeV in $pp$ collisions at 0.9 and 7~TeV.
Multiplicity dependence of the BEC parameters  revealed a saturation effect. 




\end{document}